\documentclass[12pt]{article}
\usepackage[utf8]{inputenc}
\usepackage[english]{babel}
\usepackage{amsmath, amsfonts, amssymb}
\usepackage{graphicx}
\usepackage{subfig}
\usepackage{float}
\usepackage{geometry}
\usepackage{hyperref}

\hypersetup{
	bookmarksopen=true,
	colorlinks=true,
	linkcolor=blue,
	citecolor=blue,
	urlcolor=black,	
	linktoc=all,
	pdftitle={Walking with cofee},
	pdfauthor={N. Guarin-Zapata},
	pdfkeywords={Pendulum, Parametric pumping, Coffee, Sloshing},
	pdfsubject={Project proposal},
	pdfpagemode=UseOutlines,
	pdfstartview=FitH
}

\author{Nicolás Guarín Zapata\\ nguarinz@eafit.edu.co}
\title{\textbf{Sloshing in coffee as a pumped pendulum}}

\begin{document}
\maketitle

\abstract{\footnotesize
We present a study of the dynamics of a cup of coffee while walking with it in
terms of a lumped model. The model considered is a planar (nonlinear) pendulum
with parametric and direct excitation. The position of the pivot is changed with
time, which leads to parametric excitation in the vertical direction; and
parametric/direct excitation in the horizontal one. For the former case, the
Method of Averaging is used to determine the regions of stability for two
steady-state solutions. In the vertical/horizontal excitation, we determined the
resonances of the system using the Method of Multiple Scales and show the
results computed with \emph{exact} numerical integration.}

\section*{Introduction}
Spilling coffee while walking is a common problem. The sloshing dynamics of the
coffee, while walking, can be understood in terms of an oscillator with
parametric excitation caused by the motion of the cup along the path. The main
objective of this project was to study the slosh dynamics of a cup of coffee
(while walking) in terms of a lumped mechanical model \cite{coffee}.

According to \cite{book:slosh_dynamics}, the motion of a free-liquid-surface has
three regimes (linear, weakly nonlinear, and strongly nonlinear). Depending on
the regime that we want to study the system, we can use a different
---lumped--- mechanical model to understand its dynamics. A common way
to model the sloshing of liquids in a cylindrical container is to consider it as
a pendulum, and obtain the parameters (length, mass, dissipation coefficient)
from the original system \cite{book:slosh_dynamics, kana89, coffee}. The model
considered is a pendulum with a (parametric) excitation that comes from the
\emph{human walking}, this excitation is composed of two degrees of freedom: one
in the plane (back-and-forth and lateral) and one vertical. Due to the main
frequencies in the \emph{human walking}, the model will consist of a single
pendulum\footnote{In \cite{book:slosh_dynamics}, the author use several coupled
pendulums to take into account different modes of vibration in the model.}.

We present a study of the dynamics of a cup of coffee while walking with it in
terms of a lumped model. The model considered is a planar (nonlinear) pendulum
with parametric and direct excitation. The position of the pivot is changed with
time, which leads to parametric excitation in the vertical direction; and
parametric/direct excitation in the horizontal one. For the former case, the
Method of Averaging is used to determine the regions of stability for two
steady-state solutions. In the vertical/horizontal excitation, we determined the
resonances of the system using the Method of Multiple Scales and show the
results computed with \emph{exact} numerical integration. An interesting outcome
of an analytical solution is the capability of relating the parameters of the
system to obtain some insight into the design process.\footnote{Something that
can be useful in the development of devices like the one described in this
patent \cite{patent:rucci}.}.

\section{Modeling}
A cup of coffee is, essentially, a cylindrical-like container filled with liquid
---that can be assumed to behave like water. And this liquid will present some
motion when a person walks with a cup of coffee at hand. The behavior of this
system (for linear and weakly nonlinear regimes) can be interpreted in terms of
the natural frequencies of the fluid. The natural frequencies of oscillation of
a frictionless, vorticity-free, and incompressible liquid in a cylindrical
container (a cup of coffee in this case) with a free liquid surface are given
by \cite{book:slosh_dynamics}
\[\omega_{nm}^2 = \frac{g \epsilon_{mn}}{R}\tanh\left(\epsilon_{mn}\frac{H}{R}\right)
\left[1 + \frac{\sigma}{\rho g}\left(\frac{\epsilon_{mn}}{R}\right)^2\right] \enspace ,\]
where $m=0,1,2,\cdots$ and $n=1,2,\cdots$.\footnote{In our case the surface
tension is negligible and the only relevant parameters are the geometrical ones.}
In this equation $H$ is the height of the cup, $R$ its radius, $g$ the gravity,
$\rho$ fluid density, and $\sigma$ is the surface tension. $\epsilon_{mn}$ are
the roots of the first derivative of the $m$-th order Bessel function
---\ $J'_m(\epsilon)=0$. 

One can match each of the modes of vibration of the liquid with a single degree
of freedom oscillator, e.g, a mass-spring system or a pendulum. According to
Ibrahim in \cite{book:slosh_dynamics}: 
\begin{quotation}
A realistic representation of the liquid dynamics inside closed containers can
be approximated by an equivalent mechanical system... For linear planar liquid 
otion, one can develop equivalent mechanical models in the form of a series of
mass-spring dashpot systems or a set of simple pendulums. For nonlinear sloshing
phenomena, other equivalent models such as spherical or compound pendulum may be
developed to emulate rotational and chaotic sloshing.
\end{quotation}

Since the sloshing dynamics of the coffee (while walking) can be understood in
terms of an oscillator with parametric excitation caused by the motion of the cup
along the path \cite{coffee}. A first approach is to consider the system as a
planar pendulum with parametric pumping. Taking into account the parameters that
we use parametric excitation for this model can come from: the change of length,
the motion of the point of gyration, the motion of the center of mass\footnote{This
is the case of a playground swing, where the center of mass oscillates around a
given position. When the oscillation is symmetric it can be proved to be
equivalent to the varying length pendulum \cite{book:pendulum}.}, and the
change of mass\footnote{Although this is less common.}. 

The system can be considered as a pendulum with varying length, which is a
system that has been studied before \cite{book:pendulum, sanmartin84}, and with
moving ---both, vertically and horizontally--- pivot. The parameters of the
system are: $r_0$ original length of the rod, $r$ length of the rod, $m$ mass
of the bob, $g$ gravitational acceleration, $x_0$ is a function of time that
describes the horizontal position of the pivot, and $z_0$ is a function of time
that describes the vertical position of the pivot. The present work restricts
the excitation of the system to the motion of the pivot, which leads to
\emph{direct} and \emph{parametric} excitations. \footnote{Appendix A shows the
parameters of the lumped model in terms of the original \emph{cup of coffee}.}
\begin{figure}[]
\centering
\includegraphics[height=6cm]{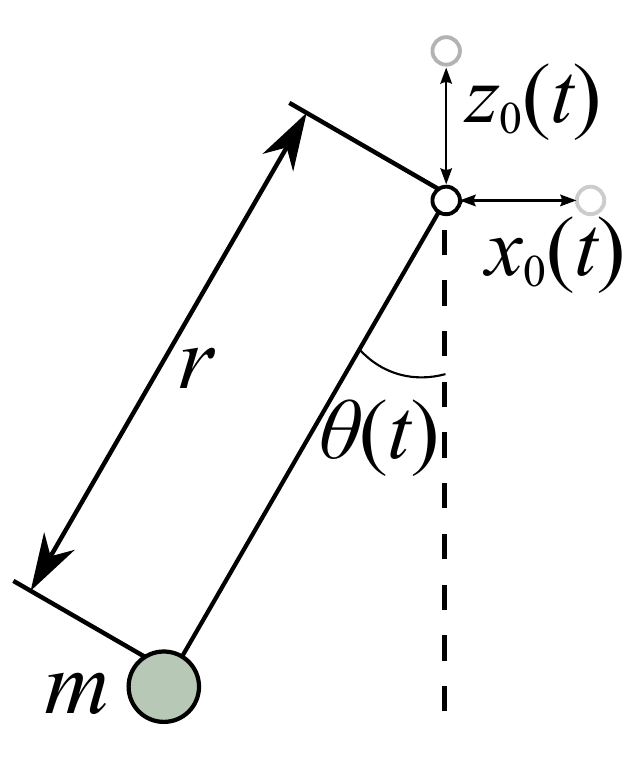}
\caption{Considered system: A pendulum with the point of gyration moving in the
plane. This motion cause direct and parametric excitation in the system.}
\label{fig:pendulum}
\end{figure}

The Lagrangian of the system is given by \cite{book:classical_mechanics}
\begin{equation}
L = T - V = \frac{m}{2}[(\dot x - \dot x_0)^2 + (\dot z + \dot z_0)^2] + mg[z + z_0]
\label{eq:lagrangian}
\end{equation}
taking
\[x = r(t) \sin\theta,\quad z = -r(t)\cos\theta\]
and using the Euler-Lagrange equation we get the differential equation, we get
\begin{equation}
r^2\ddot\theta + r[g + \ddot z_0]\sin\theta + r\ddot x_0\cos\theta + 2r\dot{r}\dot\theta = 0 \enspace .
\label{eq:ode}
\end{equation}
In the general case $x_0$, $z_0$ and $r$ are functions of time, and we can talk
about having \emph{modulated stiffness}, \emph{damping}, and \emph{inertia}.
Where the $r$ accompanies the second derivative of the angle $\theta$ and the
last term that appears due to the time dependence of the length \footnote{This
term can be compared with a damping term due to its dependence on the first
derivative of $\theta$.} \cite{book:variational_mechanics}. 
If we expand Eq. \eqref{eq:ode} in a two terms Taylor series in $\theta$ around
0, we get
\begin{equation}
r^2\ddot\theta + r[g + \ddot z_0]\left\lbrace \theta
- \frac{\theta^3}{6} \right\rbrace + r\ddot x_0\frac{\theta^2}{2}
+ 2r\dot{r}\dot\theta = -r\ddot x_0\ \enspace ,
\label{eq:ode_approx}
\end{equation}
that is a Mathieu equation with quadratic an cubic nonlinearities and a direct
excitation term.

\section{Results}
In \cite{coffee} the authors already presented an analysis based on numerical
and experimental results for the sloshing dynamics of the coffee. Here, we
emphasize our effort in approximated solutions via perturbation methods, namely:
the Method of Averaging and the Method of Multiple Scales
\cite{book:perturbation}.

\subsection{Vertical motion of the pivot: Method of Averaging}
If we consider just the vertical motion of the pivot, the (nondimensional)
differential equations turn to be
\begin{equation}
u'' + [1 + \varepsilon\lambda \Omega^2 \cos\Omega\tau]\left( u - \frac{\varepsilon^2 u^3}{6}\right)= 0 \enspace ,
\end{equation}
where
\begin{align*}
\tau =& \omega t\\
\omega_0^2 =& \frac{g}{r_0}\\
\varepsilon\lambda =& -\frac{\Delta z}{r_0}\\
\Omega =& \frac{\omega}{\omega_0} \enspace ,
\end{align*}
being $\lambda$ the ratio between the \emph{effective length} and amplitude of
oscillation, and $\varepsilon$ a bookkeeping parameter.

We can study this nonlinear differential equation using the Method of Averaging.
In Cartesian coordinates, we propose a solution of the form
\begin{align}
u(\tau) =& X(\tau)\cos\left(\frac{\Omega}{2}\tau\right) + Y(\tau)\sin\left(\frac{\Omega}{2}\tau\right)\\
u'(\tau) =& -\frac{\Omega}{2}X(\tau)\sin\left(\frac{\Omega}{2}\tau\right) + \frac{\Omega}{2}Y(\tau)\cos\left(\frac{\Omega}{2}\tau\right)
\end{align}
that is a constrained coordinate transformation. From the restriction in the
first derivative, the substitution in the differential equation and averaging
over one period we obtain two (autonomous) differential equations for $X'$ and
$Y'$ and introducing a detuning parameter ($\varepsilon\sigma = \Omega - 2$),
namely
\begin{align}
X' =& \frac{\pi \varepsilon Y}{12(\varepsilon\sigma + 2)^2}[2\lambda\varepsilon^2 Y^2
   - 3\varepsilon Y^2 - 3\varepsilon X^2 - 6\varepsilon \sigma^2 - 24\sigma - 12 \lambda]\\
Y' =& \frac{\pi \varepsilon X}{12(\varepsilon\sigma + 2)^2}[3\varepsilon Y^2
   + 2\lambda\varepsilon^2 X^2 + 3\varepsilon X^2 + 6\varepsilon \sigma^2 + 24\sigma - 12 \lambda] \enspace ,
\end{align}
we can solve this differential equation for the transient response, or equate
$(X',Y') = (0,0)$ and solve for $X$ and $Y$ to get the steady-state solution.
The solutions for the steady-state case are
\[X=Y=0\]
or
\begin{align*}
X^2 =& -\frac{3}{\lambda \varepsilon^3} (\varepsilon^2\sigma^2 + 4\varepsilon\sigma - 2\lambda\varepsilon + 6)\\
Y^2 =& \frac{3}{\lambda \varepsilon^3} (\varepsilon^2\sigma^2 + 4\varepsilon\sigma + 2\lambda\varepsilon + 6) \enspace .
\end{align*}

\begin{figure}
\centering
\includegraphics[height=8cm]{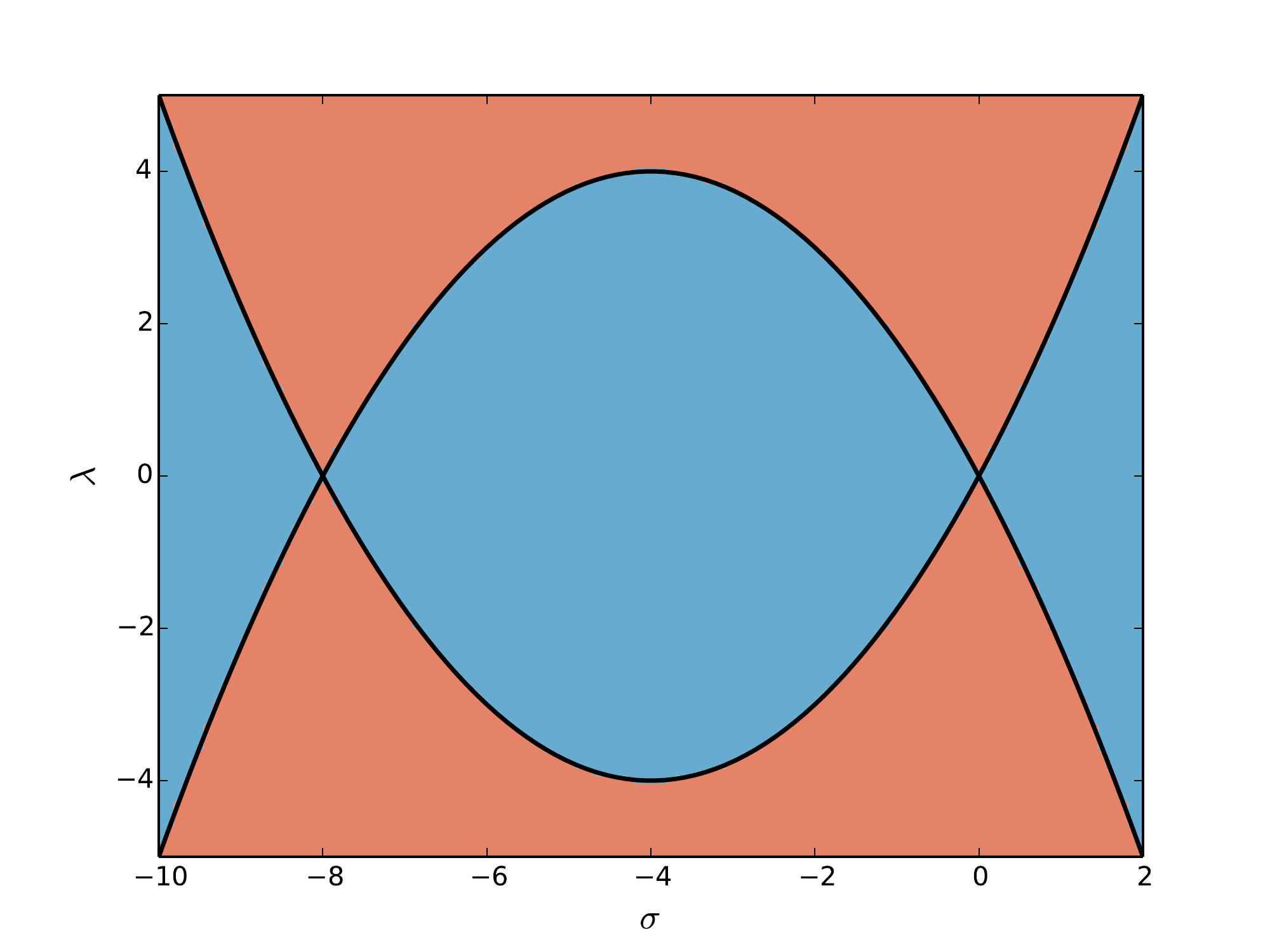} 
\caption{Stability of the system around the point (0,0). Regions in blue refer
to stable behavior, while regions in red to unstable behavior.}
\label{fig:stability_fp0}
\end{figure}

We can compute the Jacobian to obtain
\[J(X,Y) = \begin{bmatrix}-\frac{\pi \,{\varepsilon}^{2}\,X\,Y}{2\,{\left( \varepsilon\,\sigma+2\right) }^{2}} & \frac{\pi \,\varepsilon\,\left( 2\,\lambda\,{\varepsilon}^{2}\,Y^{2}-3\,\varepsilon\,Y^{2}-\varepsilon\,X^{2}-2\,\varepsilon\,\sigma^{2}-8\,\sigma-4\,\lambda\right) }{4\,{\left( \varepsilon\,\sigma+2\right) }^{2}}\cr \frac{\pi \,\varepsilon\,\left( \varepsilon\,Y^{2}+2\,\lambda\,{\varepsilon}^{2}\,X^{2}+3\,\varepsilon\,X^{2}+2\,\varepsilon\,\sigma^{2}+8\,\sigma-4\,\lambda\right) }{4\,{\left( \varepsilon\,\sigma+2\right) }^{2}} & \frac{\pi \,{\varepsilon}^{2}\,X\,Y}{2\,{\left( \varepsilon\,\sigma+2\right) }^{2}}\end{bmatrix}\]
around $(0,0)$ gives
\[J(0,0) = \frac{\pi\varepsilon}{2(\varepsilon\sigma + 2)^2}\begin{bmatrix}
0 & -(\varepsilon \sigma^2 + 4\sigma + 2\lambda) \\ 
\varepsilon \sigma^2 + 4\sigma + 2\lambda & 0
\end{bmatrix} \enspace ,\]
the stability of this fixed point is given by the determinant (since the trace
is equal to zero)
\[\Delta = \frac{\pi^2\varepsilon^2}{4(\varepsilon +2)^4}(\varepsilon\sigma^2 + 4\sigma - 2\lambda)(\varepsilon \sigma^2 + 4\sigma + 2\lambda) \enspace ,\]
this gives the critical values
\[\lambda = -\frac{1}{2}(\varepsilon\sigma_1^2 + 4\sigma_1),\quad \lambda = \frac{1}{2}(\varepsilon\sigma_2^2 + 4\sigma_2)\enspace .\]
The stability of the system is depicted in figure \ref{fig:stability_fp0}.

\begin{figure}
\centering
\includegraphics[height=8cm]{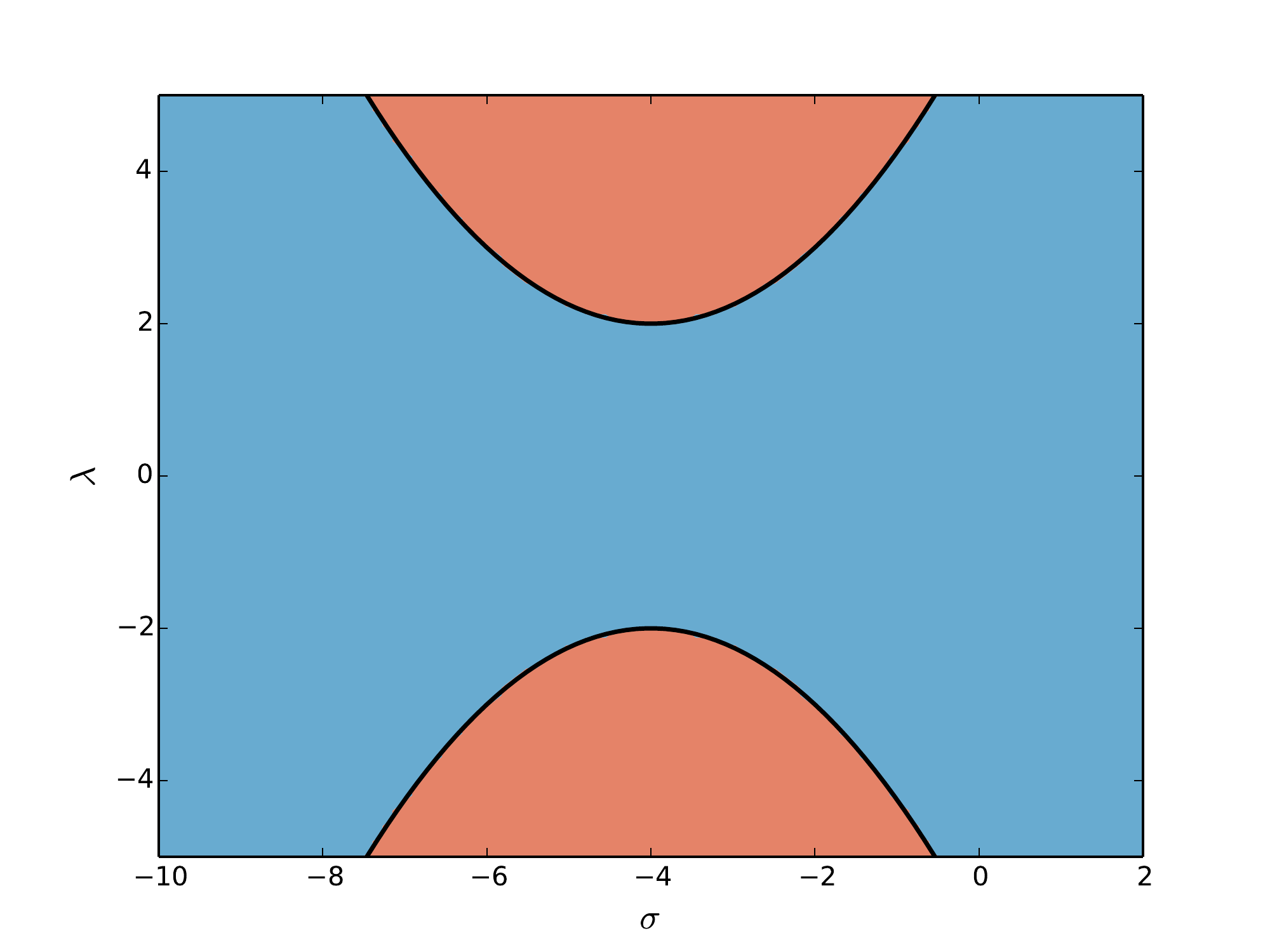} 
\caption{Stability of the system around the second fixed point. Regions in blue
refer to stable behavior, while regions in red to unstable behavior.}
\label{fig:stability_fp1}
\end{figure}

For the other fixed point the determinant is
\[\Delta = \frac{{\pi }^{2}\,\left( {\varepsilon}^{2}\,{\sigma}^{2}+4\,\varepsilon\,\sigma-2\,\lambda\,\varepsilon+6\right) \,\left( {\varepsilon}^{2}\,{\sigma}^{2}+4\,\varepsilon\,\sigma+2\,\lambda\,\varepsilon+6\right) }{{\left( \varepsilon\,\sigma+2\right) }^{4}}\]
this gives the critical values
\[\lambda = -\frac{1}{2\varepsilon}(\varepsilon^2 \sigma_1^2 + 4\varepsilon\sigma_1 + 6),\quad \lambda = \frac{1}{2\varepsilon}(\varepsilon^2\sigma_2^2 + 4\varepsilon\sigma_2 + 6)\enspace .\]
This curves present an extremum at $\sigma=-2\epsilon$. The stability of the
system is depicted in figure \ref{fig:stability_fp1}.

For parametric pumping, the optimum effect is achieved by pumping at twice the
natural frequency of the pendulum. This is constant for the linear pendulum,
due to isochrony condition. When the pendulum increases the amplitude the
period increase as
\[T = 4\sqrt{\frac{r_0}{2g}}\int\limits_{0}^{\theta_0} \frac{1}{\sqrt{\cos\theta - \cos\theta_0}}d\theta \enspace ,\]
being $\theta_0$ the initial amplitude, and the angular frequency
diminishes.\footnote{Hence, as the amplitude increases the optimal frequency of
parametric excitation decreases.} This gives us an explanation about the wider
stability for the second (non-trivial) solution. Around (0,0) the linear
analysis of stability resembles the linear case, while in the second fixed point
we are not around this point and the amplitude of the oscillation would affect
the (parametric) resonances of the system.

\subsection{Vertical and horizontal motion of the pivot: Method of Multiple Scales}
If we keep both, the vertical and horizontal motion of the pivot we get
parametric and direct excitation in the differential equation. After
normalization reads
\[u'' + [1 + \varepsilon \lambda_1\Omega^2\cos\Omega \tau]\left\lbrace u
  - \frac{\varepsilon^2 u^3}{6}\right\rbrace
  - \varepsilon^2 \lambda_2\Omega^2 \cos\Omega \tau \frac{u^2}{2}
  = - \lambda_2 \Omega^2 \cos\Omega \tau \]
with
\begin{align*}
&\tau = \omega t   {\omega_0} 
&\varepsilon\lambda_1 = -\frac{\Delta z}{r_0}\\
&\omega_0^2 = \frac{g}{r_0} &\Omega = \frac{\omega}{\omega_0}\\
&\varepsilon\lambda_1 = -\frac{\Delta z}{r_0}  & \enspace ,
\end{align*}
when, one more time $\lambda_1$ and $\lambda_2$ refers to the ratios between
the amplitude of the parametric excitations and the length of the pendulum.	

In this case, it is interesting to find the resonances that the system can
achieve, that will differ from the single resonance from the linear case. Using
the method of multiple scales we propose a solution
\[u = u_0 + \varepsilon u_1 + \varepsilon^2 u_2 +\cdots \enspace ,\]
and
\[t = T_0 + \varepsilon T_1 +  \varepsilon^2 T_2 + \cdots \enspace . \]
Replacing, and grouping by powers of $\varepsilon$ we get
\begin{align*}
\varepsilon^0: D_0^2 u_0 + u_0 =& =-\lambda_2 \Omega^2\cos\Omega T_0 \\
\varepsilon^1: D_0^2 u_1 + u_1 =& -2D_0D_1 u_0 - u_0\lambda_1 \cos\Omega T_0
  + \frac{1}{2}\lambda_2\Omega^2 u_0^2\cos\Omega T_0\\
\varepsilon^2: D_0^2 u_2 + u_2 =& -\lambda_1 u_1 \cos\Omega T_0 - 2D_0 D_1 u_1\\
     &- D_1^2 u_0 + \frac{u_0^3}{6} + \lambda_2 u_0 u_1 \Omega^2 \cos\Omega T_0
\end{align*}
Solving sequentially the set of equations, and enforcing secular terms to be
zero we get as resonances
\[\Omega \in \left\lbrace \frac{1}{2},\ 1,\ 2,\ 3 \right\rbrace\]

\begin{figure}[!h]
\centering
\subfloat[$\varepsilon=0.5$, $\lambda_1=0.2$ and $\lambda_2=0.2$.]{\includegraphics[width=0.45\textwidth]{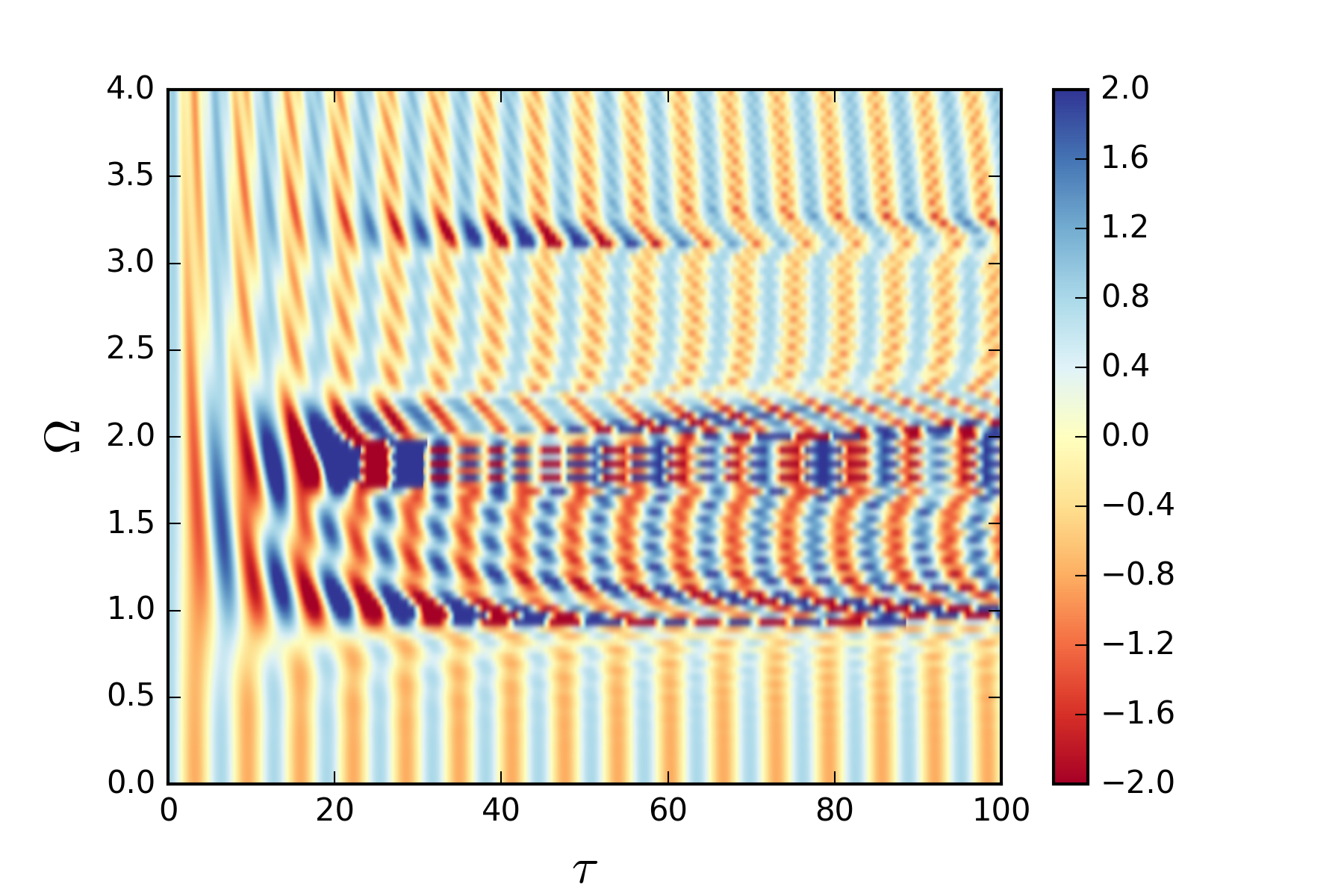}}\,
\subfloat[$\varepsilon=0.2$, $\lambda_1=0.1$ and $\lambda_2=0.3$.]{\includegraphics[width=0.45\textwidth]{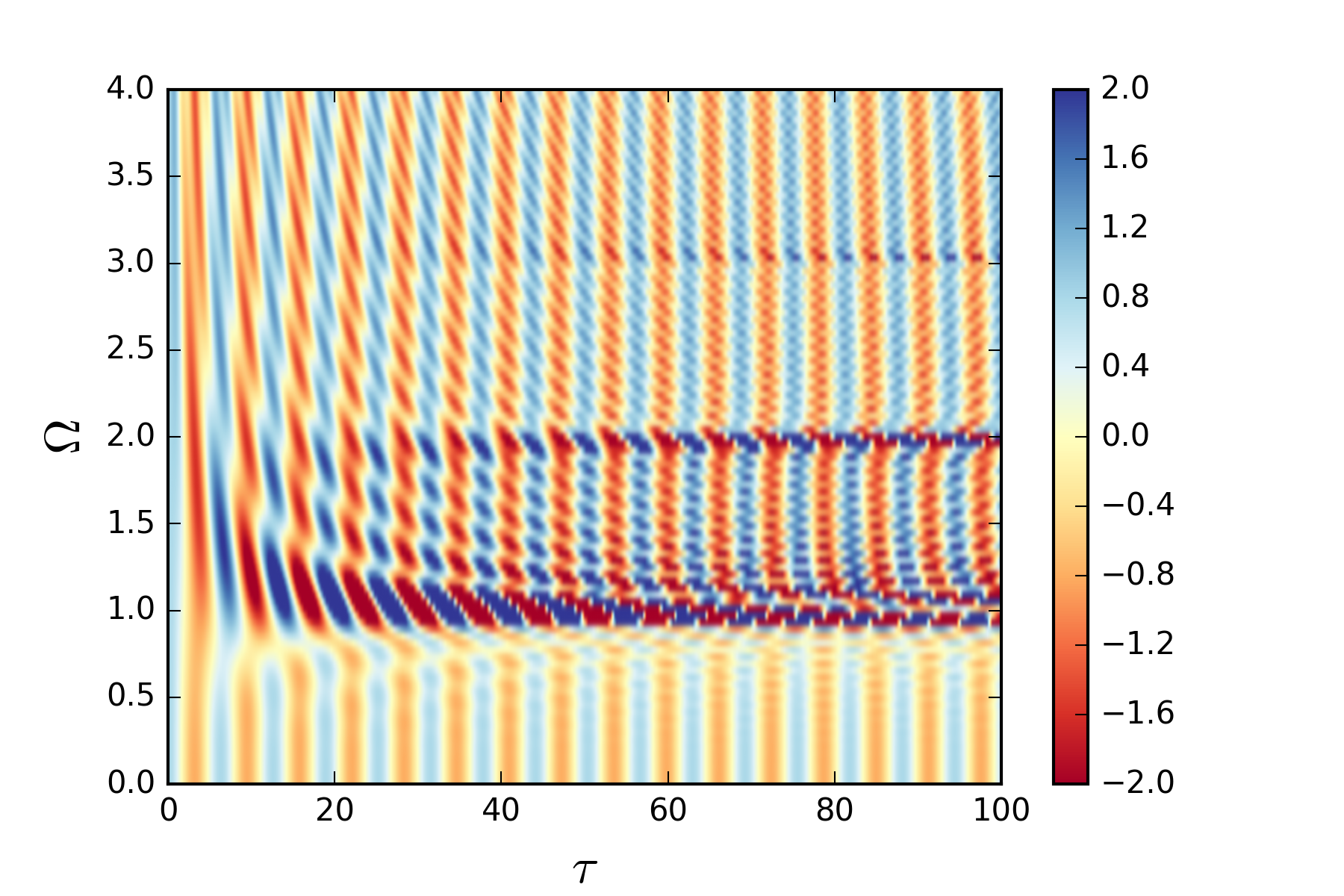}}
\caption{Time response for two sets of different parameters of the equation; in
both cases $\theta(0)=\pi/4$ and $\dot\theta(0)=0$.}
\label{fig:time_sweep}
\end{figure}

\begin{figure}[!h]
\centering
\subfloat[Solution around $\Omega=1$.]{\includegraphics[width=0.45\textwidth]{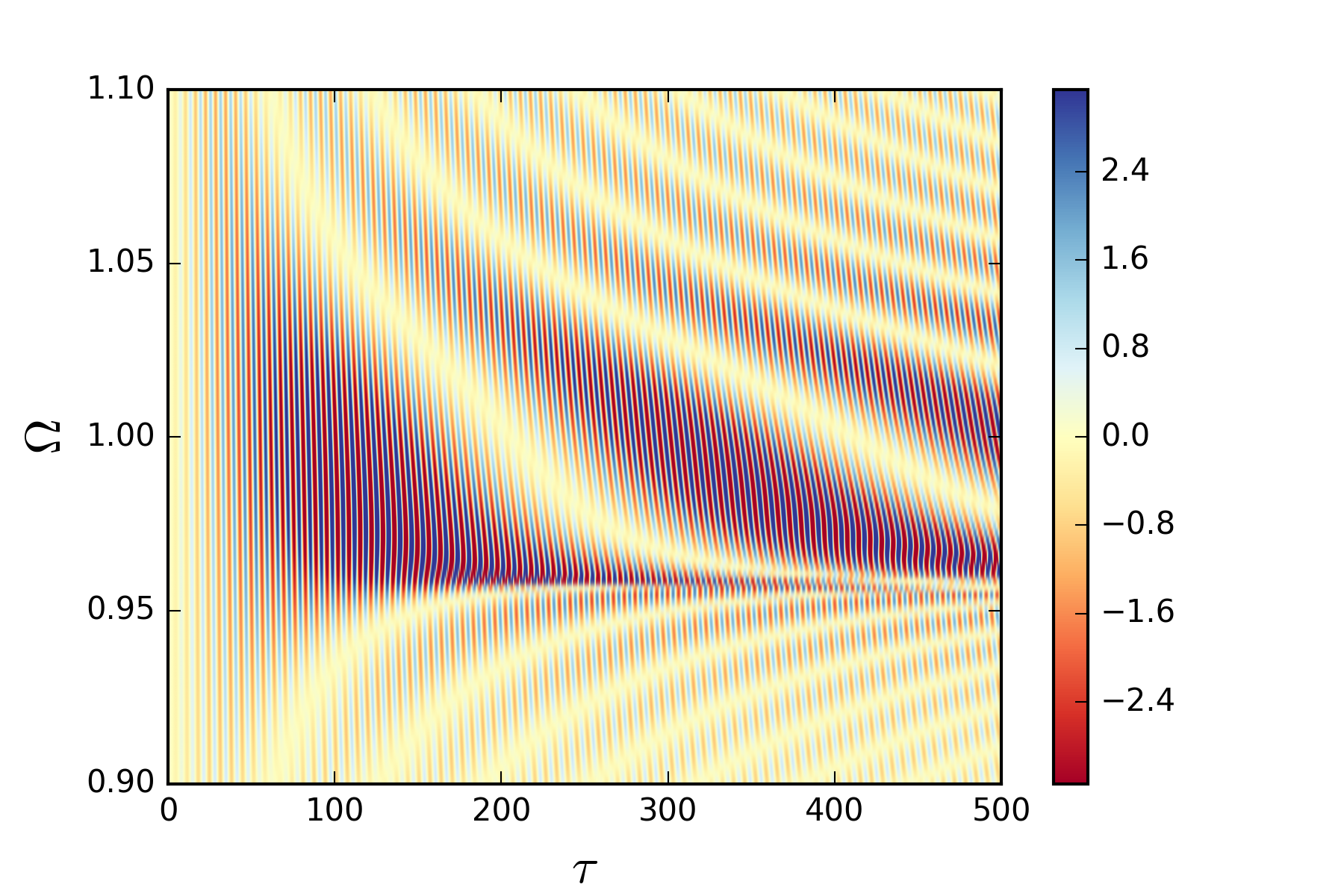}}\,
\subfloat[Solution around $\Omega=2$.]{\includegraphics[width=0.45\textwidth]{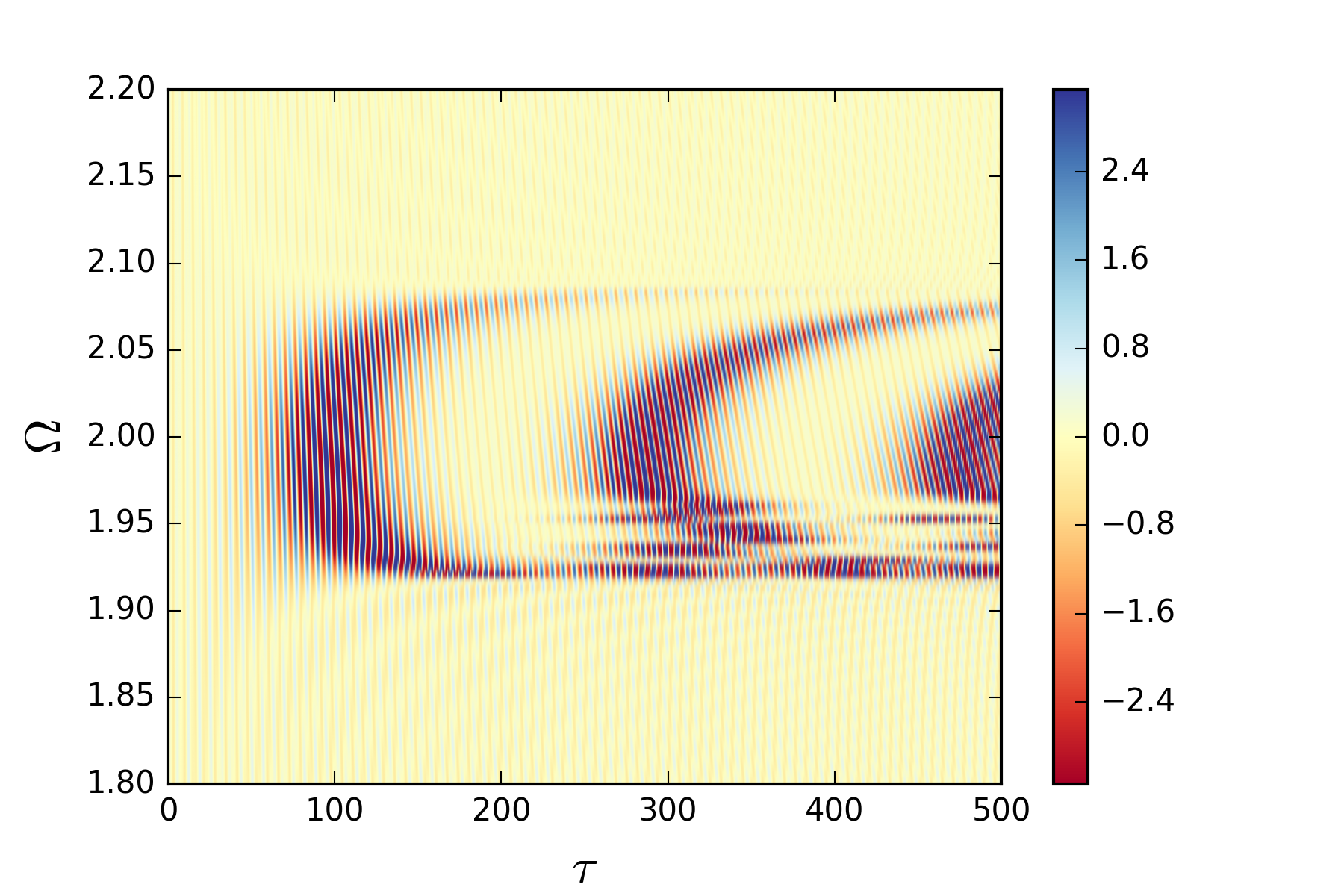}}
\caption{Time response around the primary and direct resonances. The parameters
used were: $\lambda_1=0.2$, $\lambda_2=0.1$, $\epsilon_0.2$, $\theta(0)=\pi/30$, and $\dot\theta(0)=0$.}
\label{fig:time_resonances}
\end{figure}

The appearence of three of the resonances was obtained via numerical integration
and are presented in Figure \ref{fig:time_sweep}. Figure
\ref{fig:time_resonances} show solutions for a range of frequencies around the
direct and primary parametric resonances, due to nonlinearities the higher
values are not present exactly at $\Omega = 1,\, 2$.

\subsection{Numerical Results}
We computed the numerical solution using \texttt{odeint} from
Scipy\footnote{\texttt{odeint} solve a system of ordinary differential equations
using \texttt{lsoda} from the Fortran library \texttt{odepack}.
\url{http://docs.scipy.org/doc/scipy/reference/generated/scipy.integrate.odeint.html}}.

\begin{figure}[!h]
\centering
\subfloat[$\Omega=0.00$.]{\includegraphics[width=0.45\textwidth]{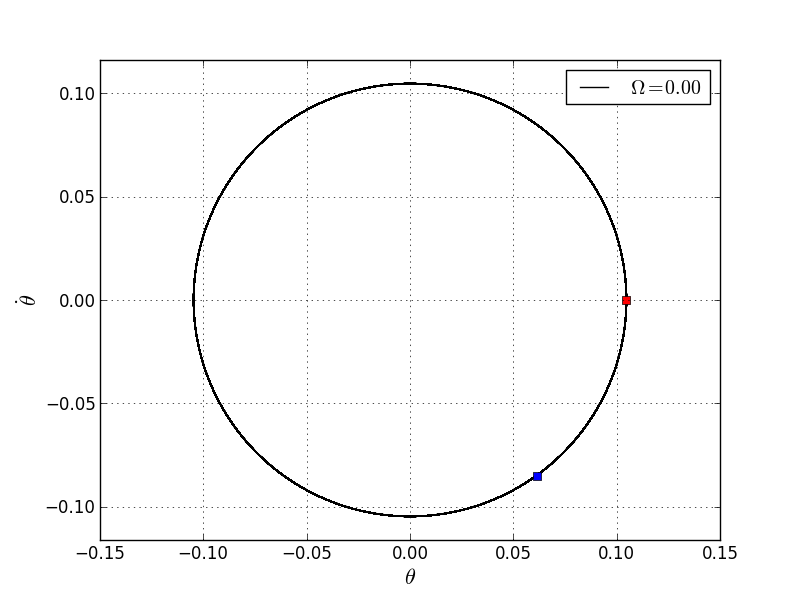}}\,
\subfloat[$\Omega=1.01$.]{\includegraphics[width=0.45\textwidth]{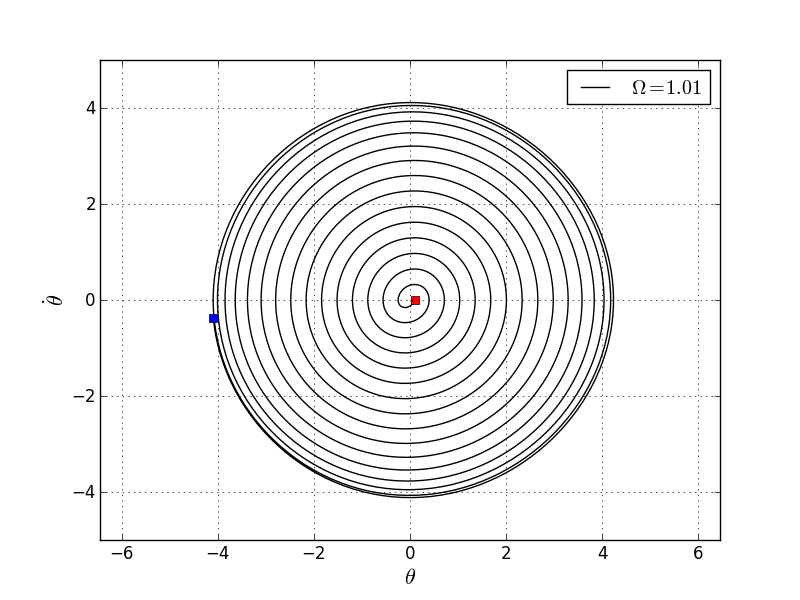}}\\
\subfloat[$\Omega=1.96$.]{\includegraphics[width=0.45\textwidth]{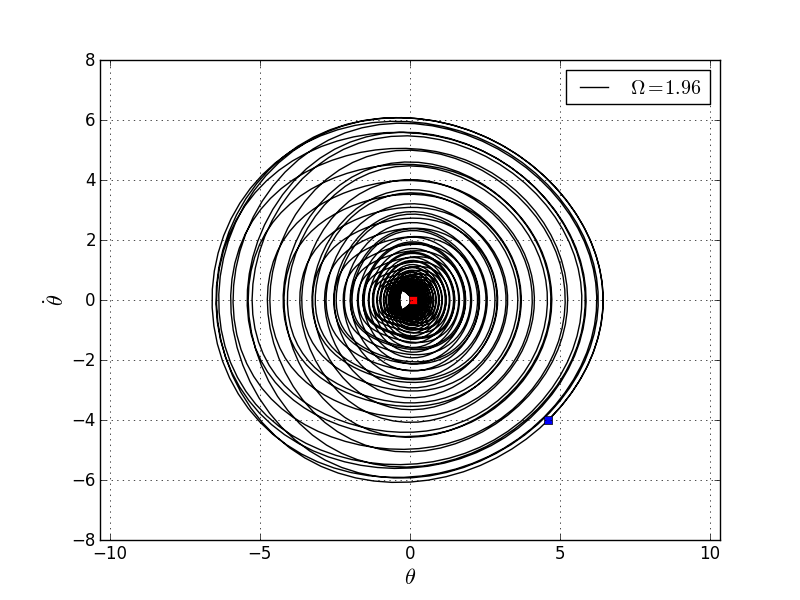}}\,
\subfloat[$\Omega=2.60$.]{\includegraphics[width=0.45\textwidth]{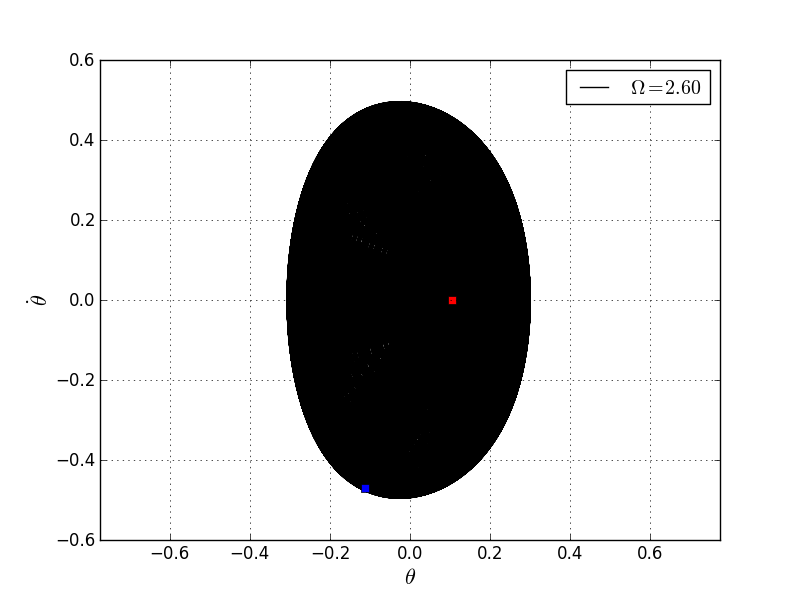}}
\end{figure}
\begin{figure}[!h]
\ContinuedFloat
\subfloat[$\Omega=2.82$.]{\includegraphics[width=0.45\textwidth]{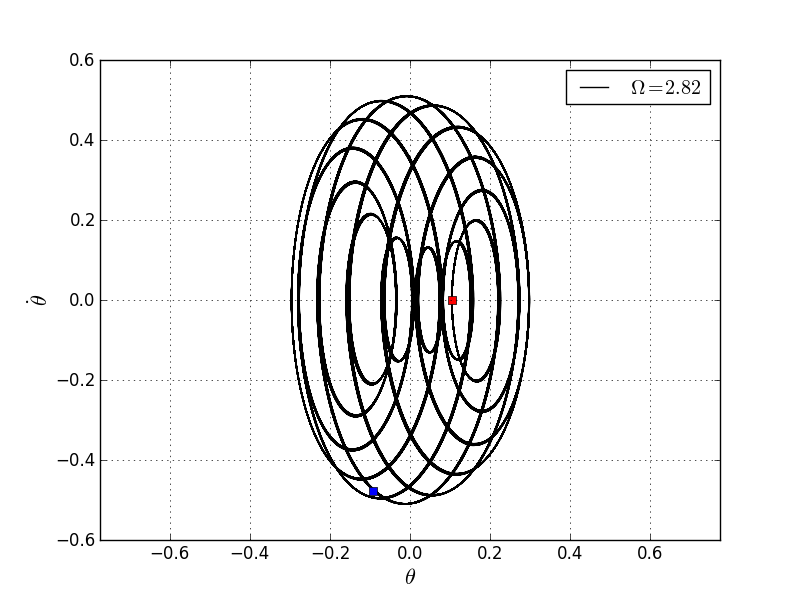}}\,
\subfloat[$\Omega=2.99$.]{\includegraphics[width=0.45\textwidth]{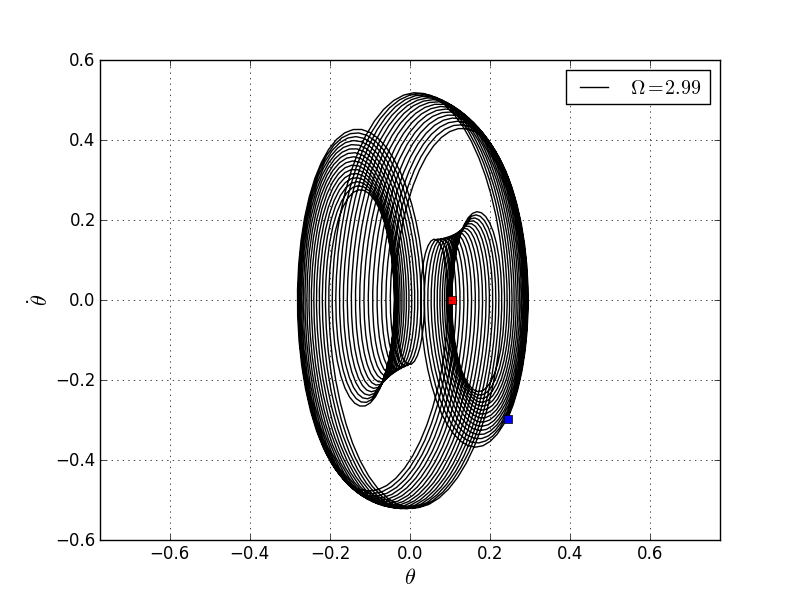}}\\
\subfloat[$\Omega=3.54$.]{\includegraphics[width=0.45\textwidth]{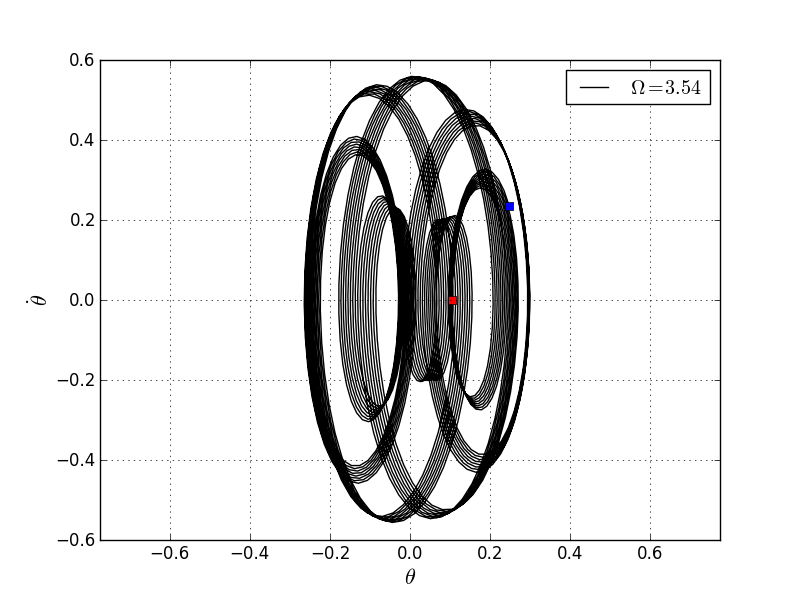}}\,
\subfloat[$\Omega=4.00$.]{\includegraphics[width=0.45\textwidth]{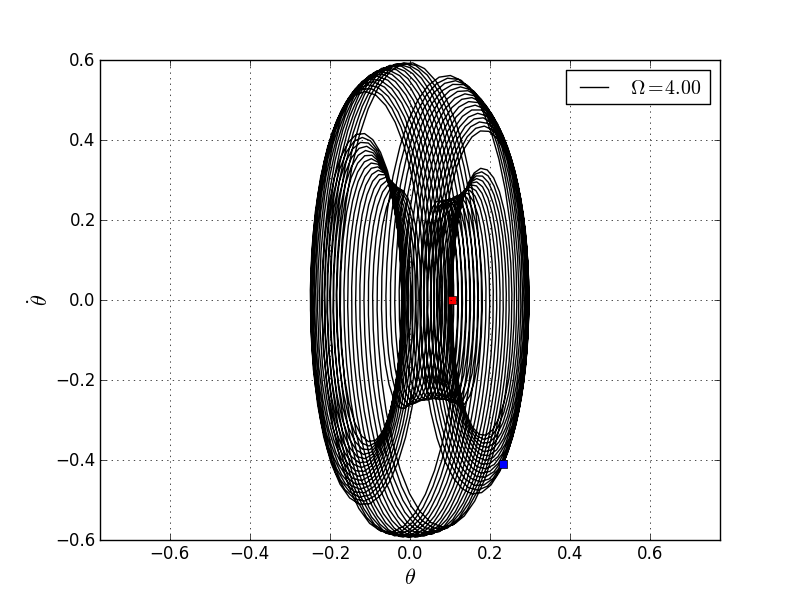}}
\caption{Phase portrait for different excitation frequencies. In all the cases
$\lambda_1=0.2$, $\lambda_2=0.1$, $\varepsilon=0.2$, $\theta(0)=0.1$ and
$\dot\theta(0)=0$. The red square is the initial point in the phase space,
while the blue one is the finishing one after $\tau=500$}
\label{fig:phase}
\end{figure}

\section{Conclusions}
A model consisting of a planar pendulum was used to model the sloshing dynamics 
of coffee. Rotating sloshing was neglected since this will need two degrees of
freedom (a spherical pendulum model).

The vertically excited pendulum presents a parametric pumping with an effective
stiffness that varies with time. This is similar to the Mathieu equation with
Duffing-like nonlinearity. When excited both, vertically and horizontally the
system presents parametric and direct excitation. Once again, the equation reads
similar to Mathieu equation with quadratic and cubic nonlinearities, which leads
to multiple resonances.

If we consider the aspect ratio of cups \emph{constant} we can have as design
parameter the radius of the cup \footnote{See Appendix A.}. Currently, the
frequency ranges for the cups and regular walking are overlapped, but we can
change this fact by \emph{tuning} the size of the cup. An increase in the size
will lead to a decrease in the natural frequency and will cease the overlapping.
From a practical point of view, this is not feasible, since one doesn't want to
have a huge cup of coffee.

Small values of $\lambda$ will translate in longer walking with coffee without
spilling, ehich can be achieved with higher sizes of cup or smaller amplitudes
of oscillation---smoother walking. The latter can be obtained \emph{focusing} in
not spilling while walking, but this can be considered as a controlled system.
This is one of the facts discussed in \cite{coffee}.

A single harmonic excitation was considered due to its main importance
\cite{coffee}, but the real walking is a complex signal with a broadband
frequency content and several resonances will appear in that case.

\section*{Appendix A: Physical Parameters}
\begin{figure}
\centering
\includegraphics[height=6cm]{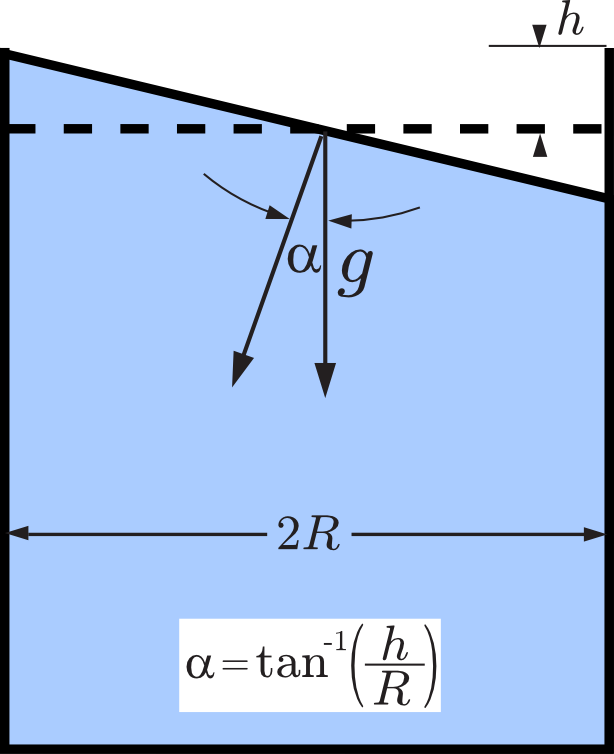} 
\caption{Geometrical description of the cup of coffee.}
\label{fig:cup}
\end{figure}

Figure \ref{fig:cup} shows a schematic of the cup of coffee. According to
\cite{coffee}, the parameters for common cups are
\begin{align*}
&R \in [2.5,\ 6.7]\, \mathrm{cm}\\
&H \in [5.7,\ 8.9]\, \mathrm{cm}\\
&h \in [5.10,\ 10]\, \mathrm{cm}\\
&\alpha \in [8^\circ,\ 16^\circ] 
\end{align*}

This combination of parameters gives a range of frequencies for the cup 
\[f_\text{cup} \in [2.6,\ 4.3]\, \mathrm{Hz} \enspace ,\]
and the frequency of walking is \cite{coffee}
\[f_\text{step} \in [1,\ 2.5]\, \mathrm{Hz} \enspace .\]

According to \cite{book:slosh_dynamics} the lumped parameters for a single
degree of freedom system are
\begin{align*}
m =& \frac{c_1 m_0}{1-c_1}\\
m_0 =& \rho \pi(H - h)R^2\\
c_1 =& \frac{R}{2.2}\tanh\left(1.84 \frac{H}{R}\right)\\
r_0 =& \frac{R}{1.84} \tanh\left(1.84 \frac{H}{R}\right) \enspace ,
\end{align*}
for simple calculation we can take $H/R\approx 1.5$, what yields
\[ r_0 \approx 0.54 R \enspace .\]

\end{document}